\documentclass[twocolumn,aps,preprintnumbers,prb,%showpacs%
,amsfonts,amsmath,amssymb,superscriptaddress,floatfix,dblfloatfix%
,10pt,lettersize]{revtex4}
\usepackage[usenames,dvipsnames]{color}
\usepackage{graphicx}
\usepackage{dcolumn}
\usepackage{bm}
\usepackage{amsmath}
\usepackage{gensymb}
\usepackage{amsfonts}
\usepackage{ragged2e}
\usepackage{psfrag}
\usepackage[pdftex,bookmarks,colorlinks=false]{hyperref}
\usepackage{booktabs}

\begin{document}

%\preprint{}

\title{Nanophotonic source of broadband quadrature squeezing}
%Nanophotonic chip source of broadband quadrature squeezing in Silicon Nitride
%Integrated source of broadband quadrature squeezing in Silicon Nitride ring resonator

\author{Robert Cernansky}
\author{Alberto Politi}
\email{A.Politi@soton.ac.uk}
\affiliation{School of Physics and Astronomy, University of Southampton, Southampton, SO17 1BJ, United Kingdom}

%\maketitle

%\date{\today}
%\keywords{Quantum optics, Nonlinear optics, integrated optics}
\begin{abstract}
%{\bf
%\noindent
Squeezed light are optical beams with variance below the Shot Noise Level. They are a key resource for quantum technologies based on photons, they can be used to achieve better precision measurements, improve security in quantum key distribution channels and as a fundamental resource for quantum computation. To date, the majority of experiments based on squeezed light have been based on non-linear crystals and discrete optical components, as the integration of quadrature squeezed states of light in a nanofabrication-friendly material %that can miniaturize optical components in a scalable architecture 
is a challenging technological task. Here we measure 0.45 dB of GHz-broad quadrature squeezing produced by a ring resonator integrated on a Silicon Nitride photonic chip that we fabricated with CMOS compatible steps. The result corrected for the off-chip losses is estimated to be 1 dB below the Shot Noise Level. %\textcolor{red}{Even if the measured level of squeezing is modest,} 
We identify and verify that the current results are limited by excess noise produced in the chip, and propose ways to reduce it. Calculations suggest that an improvement in the optical properties of the chip achievable with existing technology can %provide a squeezing level in excess of 13 dB. These results open new possibilities to integrate continuous variable sources on a photonic chip to 
develop scalable quantum technologies based on light.
\end{abstract} \maketitle

%\section{Introduction}

Squeezed states of light are a fundamental building block of quantum optics, as they are capable of generating entangled states in the continuous variable (CV) regime \cite{Weedbrook2012}. For this reason, they are the basis to demonstrate fundamental physics principles and develop quantum technologies. For example, squeezed light has been used to generate entanglement \cite{Ou1992}, as a resource required for quantum teleportation \cite{Furusawa1998}, and to produce Schrodinger cat states \cite{Ourjoumtsev2007}. In quantum key distribution, squeezed states can be employed to enhance security \cite{Gehring2015} with high compatibility with conventional optical communication technology. In sensing, squeezing has been demonstrated as the optimal resource to use in interferometers to achieve sub-Shot Noise Level (SNL) measurements \cite{Caves1981}, and its use for gravitational wave detection has demonstrated outstanding broadband sensitivity \cite{Aasi2013}. Finally, the hybrydization of CV with single photons has been recently proposed to achieve high fidelity logical operations \cite{Lee2011}, \cite{Takeda2013} for the development of quantum computers able to operate with error correction protocols \cite{Fukui2018}. The use of squeezing, rather than single photons, allows for the unconditional generation of entanglement\cite{Furusawa1998}, allowing deterministic schemes to achieve quantum advantages over the classical approach.

However, contrary to quantum optics based on single photons where the integration of photon sources \cite{Silverstone2013} circuits \cite{Politi2008} and detectors \cite{Pernice2012} has boosted the complexity of achievable experiments \cite{Wang2018},
%been an active research area in the last decade, 
limited progress has been made in integrated CV photonics. 
Simple waveguide circuits have been used for the generation of CV entanglement \cite{Masada2015}, and integrated homodyne detectors demonstrated the detection of quantum states \cite{Raffaelli2018}. The integration of these capabilities with on-chip squeezing sources can reduce the losses that degrade the quantumness of CV experiments and provide phase stability without the requirement for complex locking electronics. Moreover, full integration will provide a route to increase the complexity of CV experiments beyond what is achievable with bulk optics and can lead to the commercialization of optical quantum technologies.
Numerous demonstrations of on chip generation based on spontaneous parametric down-conversion have been achieved in periodically poled Lithium Niobate (PPLN) waveguides at telecommunication wavelength \cite{Suhara2009}. Even though recent results of on chip generation and manipulation of squeezed states are remarkable \cite{Lenzini2018},\cite{Mondain2018}, the diffusion fabrication process behind such structures results in low modal confinement, long interaction lengths and high bending losses which strongly limit the potential application for complex on-chip experiments and prevent the direct integration of either photodetectors or superconductive single photon detectors (SSPD). This is in contrast to silicon-based waveguides where the integration of efficient germanium detectors \cite{Ahn2007} and SSPD \cite{Gaggero2018} has already been demonstrated, and wafer-scale fabrication of waveguides with tight confinement and low propagation losses is well developed. This makes the demonstration of squeezing in CMOS compatible materials highly desirable.

Here, we measure 0.45 dB and infer 1 dB of broadband quadrature squeezing at telecommunication wavelengths using an on chip Silicon Nitride (SiN) microring structure that was fabricated with CMOS compatible fabrication steps. 
SiN allows low loss propagation from the UV to the Mid-IR spectrum, and absence of two-photon absorption up to visible wavelengths. Highly enhanced light-matter interactions in SiN resonators based on four wave mixing has not only been used to generate single photons at near-visible \cite{Cernansky2018} and telecommunication wavelength \cite{Reimer2014} but also for efficient creation of twin beam squeezed states \cite{Dutt2015}.
This represents an important advance for CV quantum optics, but it requires additional wavelength filtering as three wavelengths are involved, and the generation of quadrature squeezing, needed for many quantum technology schemes,  requires additional complex feedback optics and electronics. To circumvent this demanding task, we use a ring resonator to enhance the self phase modulation and generate squeezed coherent states by Kerr effect as proposed by Hoff et al. \cite{Hoff2015}. Countrary to the original theoretical scheme, we produce two counter propagating squeezed coherent states that re-interfere in an integrated Sagnac interferometer to generate a single quadrature squeezed state \cite{Shirasaki1990}. This technique has been widely investigated in experiments based on optical fibers \cite{Bergman1991}, as a way to reject spurious noise and reduce the optical power to avoid detector saturation. 

\begin{figure}[htbp]
\centering
\includegraphics[width=\linewidth]{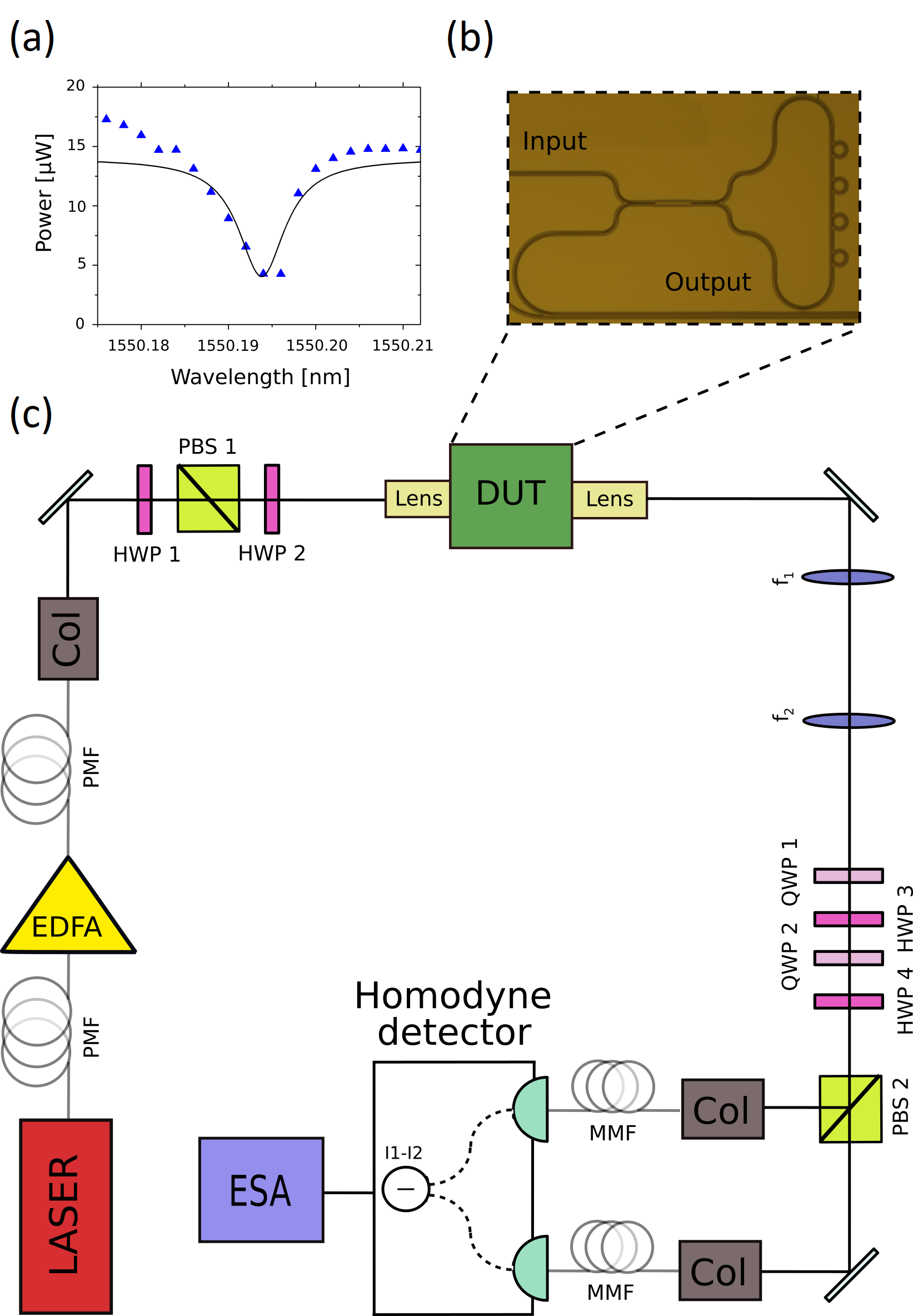}
\caption{Experimental setup (a) Transmission spectrum of the SiN chip, showing the overcoupled resonance with loaded Q-factor of 238,000 used for squeezing generation. (c) Optical image of the photonic chip. (d) Schematic setup for on chip generation of broadband quadrature squeezed states. PMF: polarization maintaining fibre, EDFA: erbium doped fibre, COL: collimator amplifier, HWP: half waveplate, QWP: quarter waveplate, PBS: polarizing beam splitter. DUT: device under test, MMF: Multi mode fibre, ESA: electronic spectrum analyser}
\label{Setup}
\end{figure}

The optical device was designed to maximize the third order non-linearity in the ring resonators (see the Methods section). We fabricate a SiN photonic circuit consisting of a 2x2 multi-mode interference (MMI) coupler with the output ports connected in a loop to form an integrated Sagnac interferometer. %The MMI is designed so that the TE mode of the waveguide reaches high visibility of the counter-propagating light beams meanwhile the visibility for the TM mode is reduced. The reason for such architecture is due to the way of encoding information of the squeezed vacuum (SV) in the horizontal (H) and the local oscillator (LO) to vertical (V) polarization, respectively. 
Inside of this loop, four microring resonators were designed in the highly overcoupled regime (escape efficiency $<70\%$). This is achieved thanks to a short single mode section that pushes the optical field out of the waveguide in order to achieve high coupling ideality\cite{Pfeiffer2017}. An optical image  of the photonic chip is reported in Fig. \ref{Setup}(a). The device was characterized by performing transmission measurements with a tunable laser. Fig. \ref{Setup}(b) displays the transmission spectrum of the overcoupled ring resonator with escape efficiency $77\%$ used for the experiment. The ring has 30 $\mu$m radius and a loaded Q-factor of 238,000, from which propagation losses of 0.32 dB/cm can be extracted.

Fig.\ref{Setup}(c) shows a schematic of the experimental setup for generation of the squeezed state (see Supplementary Information). An input beam in diagonal polarization was coupled to the photonic chip via high numerical aperture aspheric lens. In the chip, the horizontal polarization is equally split in the 50/50 MMI and coupled to the ring resonator producing two counter-propagating squeezed coherent states thanks to the self phase modulation based on the third-order nonlinear Kerr effect \cite{Bergman1991}. The beams re-interfere on the MMI and produce an attenuated quadrature squeezed state at the output port, while the majority of the pump is rejected in the input port. At the same time, vertical polarization is unequally split by the MMI (59/41 ratio); the beams counter-propagate back towards the beam splitter where they partially interfere with lower visibility, so that a mW-level beam of light co-propagates alongside the quadrature squeezed state and can be used as local oscillator. This configuration is chosen to simplify the control of the experiment, since there is high phase stability between the local oscillator and squeezed state beams as they share the same optical path. Both beams were out-coupled via an additional lens. Using four waveplates we were able to measure the noise power at different quadratures with a polarization homodyne detection scheme (see Supplementary Information).

\begin{figure}[htbp]
\centering
\includegraphics[width=0.95\linewidth]{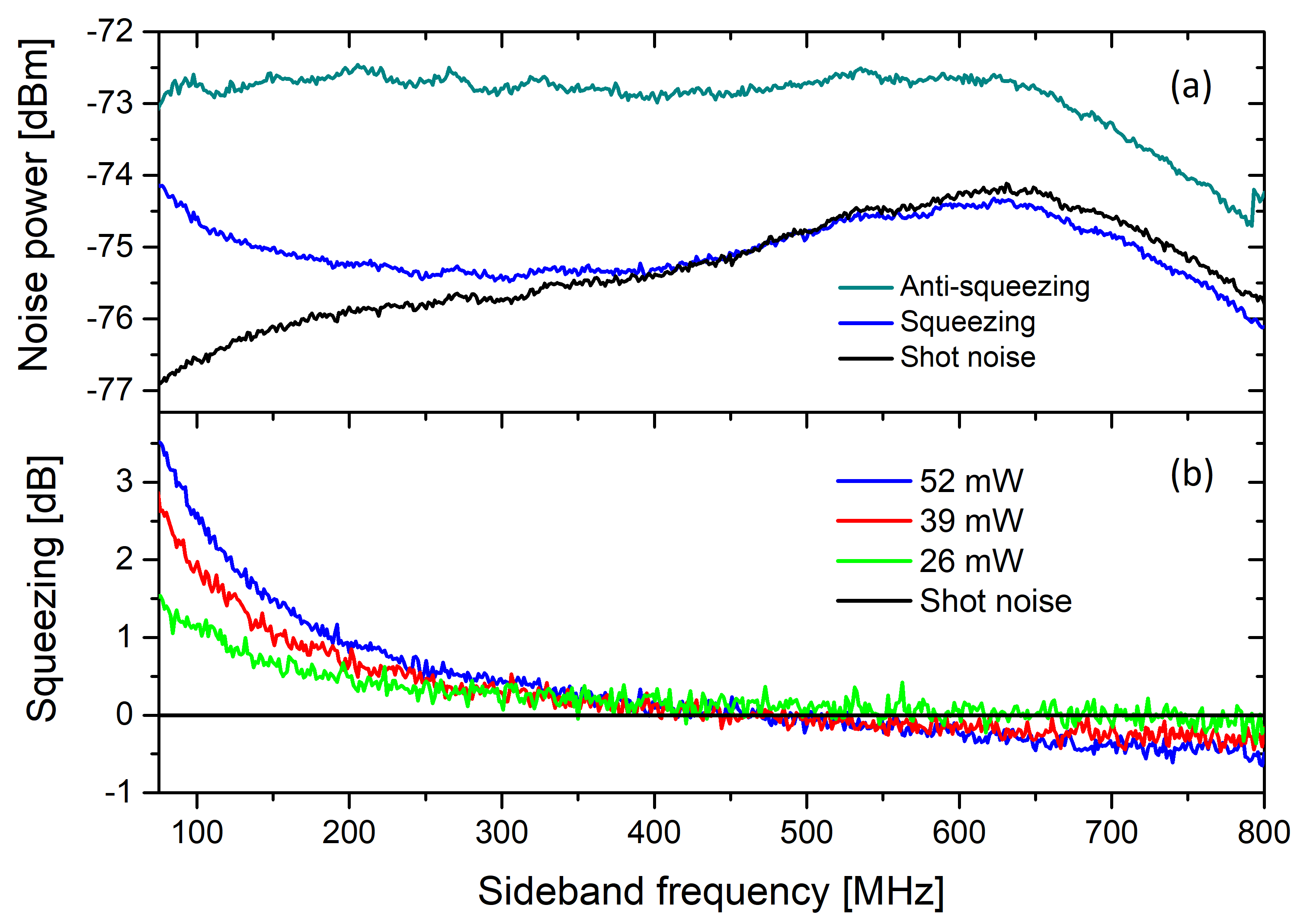}
\caption{ a) Noise spectrum for measured squeezing, anti-squeezing and shot noise  for an on-chip pump power of 52mW. The video (VBW) and radio (RBW) bandwidths are set up at 100 kHz and 20 Hz, respectively. Each line is an average of five measurements where each measurement have a sweep time of 10s. b) Squeezing spectrum, for three different input powers before the 50/50 beam splitter, that has been corrected for the noise of the detector and normalized to the shot noise level.}
\label{Squeezing1}
\end{figure}

The noise spectrum of the light collected from the chip is presented in Fig. \ref{Squeezing1}(a) for three different input powers in the input waveguide (26mW, 39mW and 52mW), while panel (b) presents the spectrum normalized to the shot noise.
Squeezing is observed spanning a frequency range of 300 MHz, with a maximum reduction of noise  of 0.45 dB. The inferred level of squeezing corrected for the measurement efficiency is estimated to be 1 dB.
The squeezing level above $\Omega \sim 800$ MHz decreases due to the response of the detector and the spectral properties of the ring resonator. This bandwidth is comparable to the one observed in down conversion parametric oscillators by monolithic cavities, but an order of magnitude greater than bow tie configurations, offering high data rate for quantum communication and cryptography while requiring modest power requirements.
No squeezing is observed at low frequencies, as excess noise above the SNL is present up to $\Omega \sim 500$ MHz. We assign the origin of this noise to the thermorefractive effect: statistical variations in the temperature of the chip drive refractive index fluctuations through the thermo-optic coefficient of the material, introducing phase noise in the propagating beam. The slow diffusion of these random temperature fluctuations results in a noise that decays with the square of the frequency\cite{Thomas2018} $\Omega^{-2}$. In the Supplementary Information we investigate the characteristics of the excess noise to experimentally confirm its thermorefractive origin and provide power, frequency and temperature dependent measurements that corroborate our model.

We further analyze the prospects to suppress this effect and to achieve high level of squeezing. The unwanted noise depends on both the thermo-optic coefficient and the thermal fluctuations induced by the local environment of the ring. 
Since for many future experiments the generation of non-Gaussian quantum states \cite{Weedbrook2012} will require the integration of SSPDs operating at cryogenic temperature, it is expected that the production of the unwanted noise will decrease.
Assuming that the ring resonator temperature is lowered to $<3$K, we expect a reduction of this noise by 50 dB, since the thermorefractive effect scales as $T^{2}$ and the thermo-optic coefficient of SiN decreases at low temperatures. In such case, we would predict a measurable squeezing level of 1.4 dB at low frequencies under the same pumping conditions. An alternative way to reduce the noise relies on a more precise control of the photonic interferometer. It has been proposed that the Sagnac works as a purifier of quantum correlations from classically correlated noisy effects such as Brillouin and Raman scattering \cite{Shirasaki1990} or even technical noise of the laser \cite{Ralph1995}. Therefore improving the contrast of our Sagnac interferometer from 23 dB to 60 dB \cite{Wilkes2016} could greatly reduce any classically correlated noise (see Supplementary Information for the effect of the Sagnac interferometer).

\begin{figure}[htbp]
\centering
\includegraphics[width=\linewidth]
{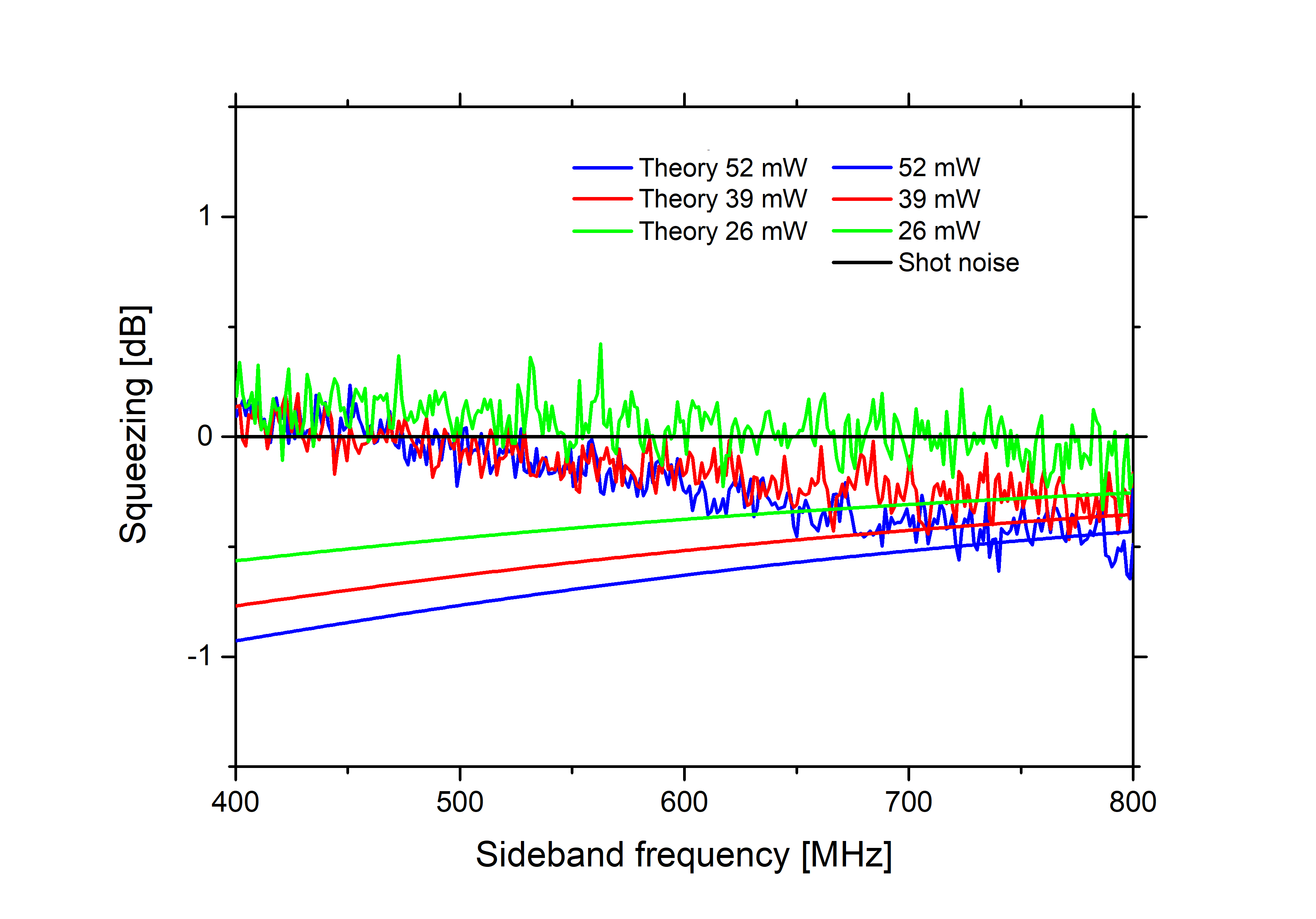}
\caption{Comparison between measured squeezed spectrum and theoretical prediction without thermorefractive noise where the calculations include the overall out-coupling losses.}
\label{Theory1}
\end{figure}

In order to understand the dynamics of generated quantum correlated states we use the theory developed by Hoff \cite{Hoff2015} for the Kerr effect in SiN microring resonators. 
We verify this model in Fig. \ref{Theory1} by comparing the measured quadrature spectra with theoretical calculations. The assumption describes reasonably well the measured squeezing at high frequencies. This can be supported by the fact that the $\Omega^{-2}$ scaling of the thermorefractive noise makes it negligible at higher frequencies, so that the noise  begins to be dominated by the Kerr squeezing.

\begin{figure}[htbp]
\centering
\includegraphics[scale=0.32]
{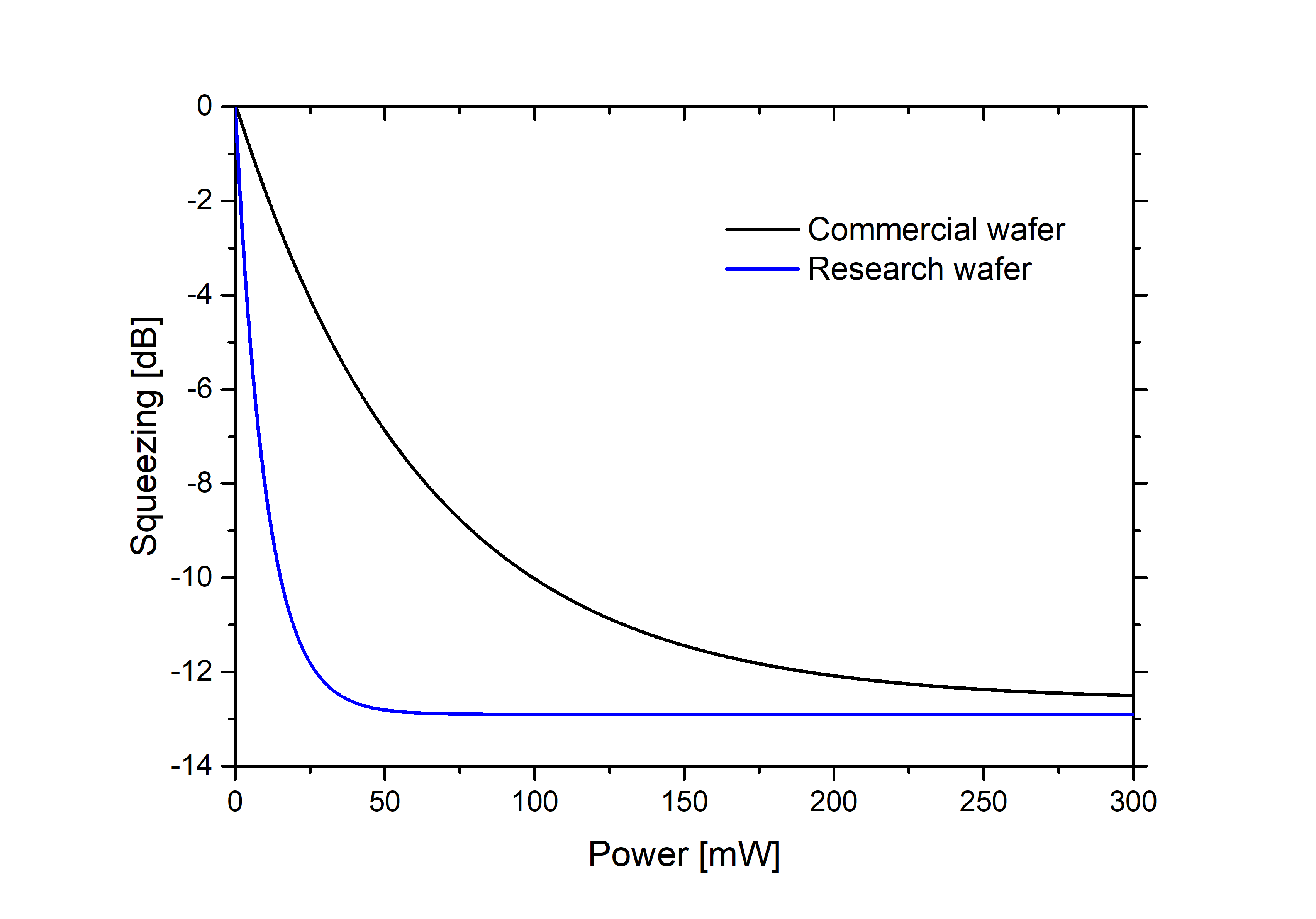}
\caption{Theoretically predicted levels of quadrature squeezing for commercially deposited SiN from \cite{Xuan2016} and research grade SiN from \cite{Ji2017}. 
The prediction is based on the waveguide and ring dimensions reported in the respective reference, and assuming an escape efficiency of 95$\%$.}
\label{Theory2}
\end{figure}

Finally, we evaluate the possible on chip generation of quadrature squeezed state with already fabricated waveguide using commercially available low pressure chemical vapour deposition, ultra low loss, SiN \cite{Xuan2016} and the highest-Q SiN ring resonators \cite{Ji2017} with an intrinsic Q factor of 13 and 37 million, respectively. In Fig.\ref{Theory2} we calculate the amount of on-chip squeezing considering a ring escape efficiency of $95 \%$. In both cases the amount of squeezing converges to 13 dB, mainly limited by the escape efficiency. More interestingly, for the higher Q factor the power required to achieve such a squeezing level is only 40mW. Such results could introduce a wide variety of future applications for continuous variable encoding for quantum computing, as 10 dB is considered to be sufficient to achieve fault tolerant universal quantum computation \cite{Fukui2018}.

%\section{Discussion}
In conclusion we have studied the generation of broadband quadrature squeezing via self phase modulation, from a CMOS-compatible SiN microring resonator in an integrated Sagnac interferometer, measured in low CW pump regime. 
Contrary to the alternative PPLN sources which are several mm long and fabricated via proton beam lithography, our process takes full advantage of the standard commercial nanofabrication techniques providing a truly scalable process. We provide demonstrations that the amount of squeezing is limited by the thermorefractive noise that is in principle present in all experiments based on self phase modulation. This noise can be heavily reduced by suppressing the temperature fluctuations operating at cryogenic temperatures and by improving the optical circuit. This would increase the measured amount of squeezing of the current sample under the same pumping conditions and provide sub-SNL light at low frequencies. The use of ultra low loss SiN would result in much higher amount of squeezing for even lower pumping powers. Furthermore, on chip entanglement can be achieved using only two ring resonators 
and simple integrated optic circuits \cite{Masada2015} providing a basis for such fundamental capabilities as quantum teleportation, cryptography and sensing. Finally, using time encoded sources of quadrature squeezed states of light, photon resolving SSPDs, delay lines and integrated germanium photodetectors, could have the potential to achieve fully integrated fault tolerant universal quantum computer %\cite{Takeda2017}, 
\cite{Alexander2018}. 
The integration of all components in a single chip would make the experiment phase-stable without the need of additional locking electronics or polarization encoding%, simplifying both the design of the photonic device and the control of the experiment. This would also imply that a CW laser and external electronics would be needed to run the experiment. 
. This will simplify both the design of the photonic device and the operation of the experiment, while removing the need for many of the elements currently required in CV experiments.
%Adding on-chip detectors will mean that just laser and electronics will be required to operate the squeezing source.
These prospects make SiN resonators excellent candidates to expand the applications of integrated sources of squeezed states of light for a broad range of future photonic quantum technology applications and go beyond the limits imposed by bulk optics.

%\section{Funding Information}

\section{Acknowledgments}
We would like to acknowledge the help of Z. Vernon and L.G. Helt for their extremely useful discussions of the origin of the low frequency technical noise and J.C. F. Matthews for helpful advice. We also acknowledge support from the Southampton Nanofabrication Centre.

This work was supported by the H2020-FETPROACT2014 Grant QUCHIP (Quantum Simulation on a Photonic Chip; grant agreement no. 641039)

Recently we become aware of the demonstration of quadrature squeezig from degenerate four wave mixing \cite{Vaidya-inprep}.

%\section{Author contributions}
%A.P. concieved the idea and supervised the project, R.C. fabricated the optical chip and performed the measurements, all authors analyzed the results, developed the theory models and wrote the paper.

%\bibliography{Bibliography}

\begin{thebibliography}{35}
\expandafter\ifx\csname natexlab\endcsname\relax\def\natexlab#1{#1}\fi
\expandafter\ifx\csname bibnamefont\endcsname\relax
  \def\bibnamefont#1{#1}\fi
\expandafter\ifx\csname bibfnamefont\endcsname\relax
  \def\bibfnamefont#1{#1}\fi
\expandafter\ifx\csname citenamefont\endcsname\relax
  \def\citenamefont#1{#1}\fi
\expandafter\ifx\csname url\endcsname\relax
  \def\url#1{\texttt{#1}}\fi
\expandafter\ifx\csname urlprefix\endcsname\relax\def\urlprefix{URL }\fi
\providecommand{\bibinfo}[2]{#2}
\providecommand{\eprint}[2][]{\url{#2}}

\bibitem[{\citenamefont{Weedbrook et~al.}(2012)\citenamefont{Weedbrook,
  Pirandola, Garc\'{\i}a-Patr\'on, Cerf, Ralph, Shapiro, and
  Lloyd}}]{Weedbrook2012}
\bibinfo{author}{\bibfnamefont{C.}~\bibnamefont{Weedbrook}},
  \bibinfo{author}{\bibfnamefont{S.}~\bibnamefont{Pirandola}},
  \bibinfo{author}{\bibfnamefont{R.}~\bibnamefont{Garc\'{\i}a-Patr\'on}},
  \bibinfo{author}{\bibfnamefont{N.~J.} \bibnamefont{Cerf}},
  \bibinfo{author}{\bibfnamefont{T.~C.} \bibnamefont{Ralph}},
  \bibinfo{author}{\bibfnamefont{J.~H.} \bibnamefont{Shapiro}},
  \bibnamefont{and} \bibinfo{author}{\bibfnamefont{S.}~\bibnamefont{Lloyd}},
  \bibinfo{journal}{Rev. Mod. Phys.} \textbf{\bibinfo{volume}{84}},
  \bibinfo{pages}{621} (\bibinfo{year}{2012}).

\bibitem[{\citenamefont{Ou et~al.}(1992)\citenamefont{Ou, Pereira, Kimble, and
  Peng}}]{Ou1992}
\bibinfo{author}{\bibfnamefont{Z.~Y.} \bibnamefont{Ou}},
  \bibinfo{author}{\bibfnamefont{S.~F.} \bibnamefont{Pereira}},
  \bibinfo{author}{\bibfnamefont{H.~J.} \bibnamefont{Kimble}},
  \bibnamefont{and} \bibinfo{author}{\bibfnamefont{K.~C.} \bibnamefont{Peng}},
  \bibinfo{journal}{Phys. Rev. Lett.} \textbf{\bibinfo{volume}{68}},
  \bibinfo{pages}{3663} (\bibinfo{year}{1992}).

\bibitem[{\citenamefont{{Furusawa} et~al.}(1998)\citenamefont{{Furusawa},
  {Sorensen}, {Braunstein}, {Fuchs}, {Kimble}, and {Polzik}}}]{Furusawa1998}
\bibinfo{author}{\bibfnamefont{A.}~\bibnamefont{{Furusawa}}},
  \bibinfo{author}{\bibfnamefont{J.~L.} \bibnamefont{{Sorensen}}},
  \bibinfo{author}{\bibfnamefont{S.~L.} \bibnamefont{{Braunstein}}},
  \bibinfo{author}{\bibfnamefont{C.~A.} \bibnamefont{{Fuchs}}},
  \bibinfo{author}{\bibfnamefont{H.~J.} \bibnamefont{{Kimble}}},
  \bibnamefont{and} \bibinfo{author}{\bibfnamefont{E.~S.}
  \bibnamefont{{Polzik}}}, \bibinfo{journal}{Science}
  \textbf{\bibinfo{volume}{282}}, \bibinfo{pages}{706} (\bibinfo{year}{1998}).

\bibitem[{\citenamefont{{Ourjoumtsev} et~al.}(2007)\citenamefont{{Ourjoumtsev},
  {Jeong}, {Tualle-Brouri}, and {Grangier}}}]{Ourjoumtsev2007}
\bibinfo{author}{\bibfnamefont{A.}~\bibnamefont{{Ourjoumtsev}}},
  \bibinfo{author}{\bibfnamefont{H.}~\bibnamefont{{Jeong}}},
  \bibinfo{author}{\bibfnamefont{R.}~\bibnamefont{{Tualle-Brouri}}},
  \bibnamefont{and}
  \bibinfo{author}{\bibfnamefont{P.}~\bibnamefont{{Grangier}}},
  \bibinfo{journal}{Nature} \textbf{\bibinfo{volume}{448}},
  \bibinfo{pages}{784} (\bibinfo{year}{2007}).

\bibitem[{\citenamefont{{Gehring} et~al.}(2015)\citenamefont{{Gehring},
  {H{\"a}ndchen}, {Duhme}, {Furrer}, {Franz}, {Pacher}, {Werner}, and
  {Schnabel}}}]{Gehring2015}
\bibinfo{author}{\bibfnamefont{T.}~\bibnamefont{{Gehring}}},
  \bibinfo{author}{\bibfnamefont{V.}~\bibnamefont{{H{\"a}ndchen}}},
  \bibinfo{author}{\bibfnamefont{J.}~\bibnamefont{{Duhme}}},
  \bibinfo{author}{\bibfnamefont{F.}~\bibnamefont{{Furrer}}},
  \bibinfo{author}{\bibfnamefont{T.}~\bibnamefont{{Franz}}},
  \bibinfo{author}{\bibfnamefont{C.}~\bibnamefont{{Pacher}}},
  \bibinfo{author}{\bibfnamefont{R.~F.} \bibnamefont{{Werner}}},
  \bibnamefont{and}
  \bibinfo{author}{\bibfnamefont{R.}~\bibnamefont{{Schnabel}}},
  \bibinfo{journal}{Nature Communications} \textbf{\bibinfo{volume}{6}},
  \bibinfo{eid}{8795} (\bibinfo{year}{2015}).

\bibitem[{\citenamefont{Caves}(1981)}]{Caves1981}
\bibinfo{author}{\bibfnamefont{C.~M.} \bibnamefont{Caves}},
  \bibinfo{journal}{Phys. Rev. D} \textbf{\bibinfo{volume}{23}},
  \bibinfo{pages}{1693} (\bibinfo{year}{1981}).

\bibitem[{\citenamefont{{Aasi} et~al.}(2013)\citenamefont{{Aasi}, {Abadie},
  {Abbott}, {Abbott}, {Abbott}, {Abernathy}, {Adams}, {Adams}, {Addesso},
  {Adhikari} et~al.}}]{Aasi2013}
\bibinfo{author}{\bibfnamefont{J.}~\bibnamefont{{Aasi}}},
  \bibinfo{author}{\bibfnamefont{J.}~\bibnamefont{{Abadie}}},
  \bibinfo{author}{\bibfnamefont{B.~P.} \bibnamefont{{Abbott}}},
  \bibinfo{author}{\bibfnamefont{R.}~\bibnamefont{{Abbott}}},
  \bibinfo{author}{\bibfnamefont{T.~D.} \bibnamefont{{Abbott}}},
  \bibinfo{author}{\bibfnamefont{M.~R.} \bibnamefont{{Abernathy}}},
  \bibinfo{author}{\bibfnamefont{C.}~\bibnamefont{{Adams}}},
  \bibinfo{author}{\bibfnamefont{T.}~\bibnamefont{{Adams}}},
  \bibinfo{author}{\bibfnamefont{P.}~\bibnamefont{{Addesso}}},
  \bibinfo{author}{\bibfnamefont{R.~X.} \bibnamefont{{Adhikari}}},
  \bibnamefont{et~al.}, \bibinfo{journal}{Nature Photonics}
  \textbf{\bibinfo{volume}{7}}, \bibinfo{pages}{613} (\bibinfo{year}{2013}).

\bibitem[{\citenamefont{{Lee} et~al.}(2011)\citenamefont{{Lee}, {Benichi},
  {Takeno}, {Takeda}, {Webb}, {Huntington}, and {Furusawa}}}]{Lee2011}
\bibinfo{author}{\bibfnamefont{N.}~\bibnamefont{{Lee}}},
  \bibinfo{author}{\bibfnamefont{H.}~\bibnamefont{{Benichi}}},
  \bibinfo{author}{\bibfnamefont{Y.}~\bibnamefont{{Takeno}}},
  \bibinfo{author}{\bibfnamefont{S.}~\bibnamefont{{Takeda}}},
  \bibinfo{author}{\bibfnamefont{J.}~\bibnamefont{{Webb}}},
  \bibinfo{author}{\bibfnamefont{E.}~\bibnamefont{{Huntington}}},
  \bibnamefont{and}
  \bibinfo{author}{\bibfnamefont{A.}~\bibnamefont{{Furusawa}}},
  \bibinfo{journal}{Science} \textbf{\bibinfo{volume}{332}},
  \bibinfo{pages}{330} (\bibinfo{year}{2011}).

\bibitem[{\citenamefont{{Takeda} et~al.}(2013)\citenamefont{{Takeda}, {Mizuta},
  {Fuwa}, {van Loock}, and {Furusawa}}}]{Takeda2013}
\bibinfo{author}{\bibfnamefont{S.}~\bibnamefont{{Takeda}}},
  \bibinfo{author}{\bibfnamefont{T.}~\bibnamefont{{Mizuta}}},
  \bibinfo{author}{\bibfnamefont{M.}~\bibnamefont{{Fuwa}}},
  \bibinfo{author}{\bibfnamefont{P.}~\bibnamefont{{van Loock}}},
  \bibnamefont{and}
  \bibinfo{author}{\bibfnamefont{A.}~\bibnamefont{{Furusawa}}},
  \bibinfo{journal}{\nat} \textbf{\bibinfo{volume}{500}}, \bibinfo{pages}{315}
  (\bibinfo{year}{2013}).

\bibitem[{\citenamefont{Fukui et~al.}(2018)\citenamefont{Fukui, Tomita,
  Okamoto, and Fujii}}]{Fukui2018}
\bibinfo{author}{\bibfnamefont{K.}~\bibnamefont{Fukui}},
  \bibinfo{author}{\bibfnamefont{A.}~\bibnamefont{Tomita}},
  \bibinfo{author}{\bibfnamefont{A.}~\bibnamefont{Okamoto}}, \bibnamefont{and}
  \bibinfo{author}{\bibfnamefont{K.}~\bibnamefont{Fujii}},
  \bibinfo{journal}{Phys. Rev. X} \textbf{\bibinfo{volume}{8}},
  \bibinfo{pages}{021054} (\bibinfo{year}{2018}).

\bibitem[{\citenamefont{{Silverstone} et~al.}(2014)\citenamefont{{Silverstone},
  {Bonneau}, {Ohira}, {Suzuki}, {Yoshida}, {Iizuka}, {Ezaki}, {Natarajan},
  {Tanner}, {Hadfield} et~al.}}]{Silverstone2013}
\bibinfo{author}{\bibfnamefont{J.~W.} \bibnamefont{{Silverstone}}},
  \bibinfo{author}{\bibfnamefont{D.}~\bibnamefont{{Bonneau}}},
  \bibinfo{author}{\bibfnamefont{K.}~\bibnamefont{{Ohira}}},
  \bibinfo{author}{\bibfnamefont{N.}~\bibnamefont{{Suzuki}}},
  \bibinfo{author}{\bibfnamefont{H.}~\bibnamefont{{Yoshida}}},
  \bibinfo{author}{\bibfnamefont{N.}~\bibnamefont{{Iizuka}}},
  \bibinfo{author}{\bibfnamefont{M.}~\bibnamefont{{Ezaki}}},
  \bibinfo{author}{\bibfnamefont{C.~M.} \bibnamefont{{Natarajan}}},
  \bibinfo{author}{\bibfnamefont{M.~G.} \bibnamefont{{Tanner}}},
  \bibinfo{author}{\bibfnamefont{R.~H.} \bibnamefont{{Hadfield}}},
  \bibnamefont{et~al.}, \bibinfo{journal}{Nature Photonics}
  \textbf{\bibinfo{volume}{8}}, \bibinfo{pages}{104} (\bibinfo{year}{2014}).

\bibitem[{\citenamefont{{Politi} et~al.}(2008)\citenamefont{{Politi}, {Cryan},
  {Rarity}, {Yu}, and {O'Brien}}}]{Politi2008}
\bibinfo{author}{\bibfnamefont{A.}~\bibnamefont{{Politi}}},
  \bibinfo{author}{\bibfnamefont{M.~J.} \bibnamefont{{Cryan}}},
  \bibinfo{author}{\bibfnamefont{J.~G.} \bibnamefont{{Rarity}}},
  \bibinfo{author}{\bibfnamefont{S.}~\bibnamefont{{Yu}}}, \bibnamefont{and}
  \bibinfo{author}{\bibfnamefont{J.~L.} \bibnamefont{{O'Brien}}},
  \bibinfo{journal}{Science} \textbf{\bibinfo{volume}{320}},
  \bibinfo{pages}{646} (\bibinfo{year}{2008}).

\bibitem[{\citenamefont{{Pernice} et~al.}(2012)\citenamefont{{Pernice},
  {Schuck}, {Minaeva}, {Li}, {Goltsman}, {Sergienko}, and
  {Tang}}}]{Pernice2012}
\bibinfo{author}{\bibfnamefont{W.~H.~P.} \bibnamefont{{Pernice}}},
  \bibinfo{author}{\bibfnamefont{C.}~\bibnamefont{{Schuck}}},
  \bibinfo{author}{\bibfnamefont{O.}~\bibnamefont{{Minaeva}}},
  \bibinfo{author}{\bibfnamefont{M.}~\bibnamefont{{Li}}},
  \bibinfo{author}{\bibfnamefont{G.~N.} \bibnamefont{{Goltsman}}},
  \bibinfo{author}{\bibfnamefont{A.~V.} \bibnamefont{{Sergienko}}},
  \bibnamefont{and} \bibinfo{author}{\bibfnamefont{H.~X.}
  \bibnamefont{{Tang}}}, \bibinfo{journal}{Nature Communications}
  \textbf{\bibinfo{volume}{3}}, \bibinfo{eid}{1325} (\bibinfo{year}{2012}).

\bibitem[{\citenamefont{Wang et~al.}(2018)\citenamefont{Wang, Paesani, Ding,
  Santagati, Skrzypczyk, Salavrakos, Tura, Augusiak, Mancinska, Bacco
  et~al.}}]{Wang2018}
\bibinfo{author}{\bibfnamefont{J.}~\bibnamefont{Wang}},
  \bibinfo{author}{\bibfnamefont{S.}~\bibnamefont{Paesani}},
  \bibinfo{author}{\bibfnamefont{Y.}~\bibnamefont{Ding}},
  \bibinfo{author}{\bibfnamefont{R.}~\bibnamefont{Santagati}},
  \bibinfo{author}{\bibfnamefont{P.}~\bibnamefont{Skrzypczyk}},
  \bibinfo{author}{\bibfnamefont{A.}~\bibnamefont{Salavrakos}},
  \bibinfo{author}{\bibfnamefont{J.}~\bibnamefont{Tura}},
  \bibinfo{author}{\bibfnamefont{R.}~\bibnamefont{Augusiak}},
  \bibinfo{author}{\bibfnamefont{L.}~\bibnamefont{Mancinska}},
  \bibinfo{author}{\bibfnamefont{D.}~\bibnamefont{Bacco}},
  \bibnamefont{et~al.}, \bibinfo{journal}{Science}
  \textbf{\bibinfo{volume}{360}}, \bibinfo{pages}{285} (\bibinfo{year}{2018}).

\bibitem[{\citenamefont{{Masada} et~al.}(2015)\citenamefont{{Masada}, {Miyata},
  {Politi}, {Hashimoto}, {O'Brien}, and {Furusawa}}}]{Masada2015}
\bibinfo{author}{\bibfnamefont{G.}~\bibnamefont{{Masada}}},
  \bibinfo{author}{\bibfnamefont{K.}~\bibnamefont{{Miyata}}},
  \bibinfo{author}{\bibfnamefont{A.}~\bibnamefont{{Politi}}},
  \bibinfo{author}{\bibfnamefont{T.}~\bibnamefont{{Hashimoto}}},
  \bibinfo{author}{\bibfnamefont{J.~L.} \bibnamefont{{O'Brien}}},
  \bibnamefont{and}
  \bibinfo{author}{\bibfnamefont{A.}~\bibnamefont{{Furusawa}}},
  \bibinfo{journal}{Nature Photonics} \textbf{\bibinfo{volume}{9}},
  \bibinfo{pages}{316} (\bibinfo{year}{2015}).

\bibitem[{\citenamefont{Raffaelli et~al.}(2018)\citenamefont{Raffaelli,
  Ferranti, Mahler, Sibson, Kennard, Santamato, Sinclair, Bonneau, Thompson,
  and Matthews}}]{Raffaelli2018}
\bibinfo{author}{\bibfnamefont{F.}~\bibnamefont{Raffaelli}},
  \bibinfo{author}{\bibfnamefont{G.}~\bibnamefont{Ferranti}},
  \bibinfo{author}{\bibfnamefont{D.~H.} \bibnamefont{Mahler}},
  \bibinfo{author}{\bibfnamefont{P.}~\bibnamefont{Sibson}},
  \bibinfo{author}{\bibfnamefont{J.~E.} \bibnamefont{Kennard}},
  \bibinfo{author}{\bibfnamefont{A.}~\bibnamefont{Santamato}},
  \bibinfo{author}{\bibfnamefont{G.}~\bibnamefont{Sinclair}},
  \bibinfo{author}{\bibfnamefont{D.}~\bibnamefont{Bonneau}},
  \bibinfo{author}{\bibfnamefont{M.~G.} \bibnamefont{Thompson}},
  \bibnamefont{and} \bibinfo{author}{\bibfnamefont{J.~C.~F.}
  \bibnamefont{Matthews}}, \bibinfo{journal}{Quantum Science and Technology}
  \textbf{\bibinfo{volume}{3}}, \bibinfo{pages}{025003} (\bibinfo{year}{2018}).

\bibitem[{\citenamefont{{Suhara}}(2009)}]{Suhara2009}
\bibinfo{author}{\bibfnamefont{T.}~\bibnamefont{{Suhara}}},
  \bibinfo{journal}{Laser \& Photonics Review} \textbf{\bibinfo{volume}{3}},
  \bibinfo{pages}{370} (\bibinfo{year}{2009}).

\bibitem[{\citenamefont{Lenzini et~al.}(2018)\citenamefont{Lenzini, Janousek,
  Thearle, Villa, Haylock, Kasture, Cui, Phan, Dao, Yonezawa
  et~al.}}]{Lenzini2018}
\bibinfo{author}{\bibfnamefont{F.}~\bibnamefont{Lenzini}},
  \bibinfo{author}{\bibfnamefont{J.}~\bibnamefont{Janousek}},
  \bibinfo{author}{\bibfnamefont{O.}~\bibnamefont{Thearle}},
  \bibinfo{author}{\bibfnamefont{M.}~\bibnamefont{Villa}},
  \bibinfo{author}{\bibfnamefont{B.}~\bibnamefont{Haylock}},
  \bibinfo{author}{\bibfnamefont{S.}~\bibnamefont{Kasture}},
  \bibinfo{author}{\bibfnamefont{L.}~\bibnamefont{Cui}},
  \bibinfo{author}{\bibfnamefont{H.-P.} \bibnamefont{Phan}},
  \bibinfo{author}{\bibfnamefont{D.~V.} \bibnamefont{Dao}},
  \bibinfo{author}{\bibfnamefont{H.}~\bibnamefont{Yonezawa}},
  \bibnamefont{et~al.}, \bibinfo{journal}{Science Advances}
  \textbf{\bibinfo{volume}{4}} (\bibinfo{year}{2018}).

\bibitem[{\citenamefont{{Mondain} et~al.}(2018)\citenamefont{{Mondain},
  {Lunghi}, {Zavatta}, {Gouzien}, {Doutre}, {De Micheli}, {Tanzilli}, and
  {D'Auria}}}]{Mondain2018}
\bibinfo{author}{\bibfnamefont{F.}~\bibnamefont{{Mondain}}},
  \bibinfo{author}{\bibfnamefont{T.}~\bibnamefont{{Lunghi}}},
  \bibinfo{author}{\bibfnamefont{A.}~\bibnamefont{{Zavatta}}},
  \bibinfo{author}{\bibfnamefont{{\'E}.}~\bibnamefont{{Gouzien}}},
  \bibinfo{author}{\bibfnamefont{F.}~\bibnamefont{{Doutre}}},
  \bibinfo{author}{\bibfnamefont{M.}~\bibnamefont{{De Micheli}}},
  \bibinfo{author}{\bibfnamefont{S.}~\bibnamefont{{Tanzilli}}},
  \bibnamefont{and}
  \bibinfo{author}{\bibfnamefont{V.}~\bibnamefont{{D'Auria}}},
  \bibinfo{journal}{arXiv e-prints}  (\bibinfo{year}{2018}),
  \eprint{1811.02097}.

\bibitem[{\citenamefont{{Ahn} et~al.}(2007)\citenamefont{{Ahn}, {Hong}, {Liu},
  {Giziewicz}, {Beals}, {Kimerling}, {Michel}, {Chen}, and
  {K{\"a}rtner}}}]{Ahn2007}
\bibinfo{author}{\bibfnamefont{D.}~\bibnamefont{{Ahn}}},
  \bibinfo{author}{\bibfnamefont{C.-Y.} \bibnamefont{{Hong}}},
  \bibinfo{author}{\bibfnamefont{J.}~\bibnamefont{{Liu}}},
  \bibinfo{author}{\bibfnamefont{W.}~\bibnamefont{{Giziewicz}}},
  \bibinfo{author}{\bibfnamefont{M.}~\bibnamefont{{Beals}}},
  \bibinfo{author}{\bibfnamefont{L.~C.} \bibnamefont{{Kimerling}}},
  \bibinfo{author}{\bibfnamefont{J.}~\bibnamefont{{Michel}}},
  \bibinfo{author}{\bibfnamefont{J.}~\bibnamefont{{Chen}}}, \bibnamefont{and}
  \bibinfo{author}{\bibfnamefont{F.~X.} \bibnamefont{{K{\"a}rtner}}},
  \bibinfo{journal}{Optics Express} \textbf{\bibinfo{volume}{15}},
  \bibinfo{pages}{3916} (\bibinfo{year}{2007}).

\bibitem[{\citenamefont{{Gaggero} et~al.}(2018)\citenamefont{{Gaggero},
  {Martini}, {Mattioli}, {Chiarello}, {Cernansky}, {Politi}, and
  {Leoni}}}]{Gaggero2018}
\bibinfo{author}{\bibfnamefont{A.}~\bibnamefont{{Gaggero}}},
  \bibinfo{author}{\bibfnamefont{F.}~\bibnamefont{{Martini}}},
  \bibinfo{author}{\bibfnamefont{F.}~\bibnamefont{{Mattioli}}},
  \bibinfo{author}{\bibfnamefont{F.}~\bibnamefont{{Chiarello}}},
  \bibinfo{author}{\bibfnamefont{R.}~\bibnamefont{{Cernansky}}},
  \bibinfo{author}{\bibfnamefont{A.}~\bibnamefont{{Politi}}}, \bibnamefont{and}
  \bibinfo{author}{\bibfnamefont{R.}~\bibnamefont{{Leoni}}},
  \bibinfo{journal}{arXiv e-prints}  (\bibinfo{year}{2018}),
  \eprint{1811.12306}.

\bibitem[{\citenamefont{{Cernansky} et~al.}(2018)\citenamefont{{Cernansky},
  {Martini}, and {Politi}}}]{Cernansky2018}
\bibinfo{author}{\bibfnamefont{R.}~\bibnamefont{{Cernansky}}},
  \bibinfo{author}{\bibfnamefont{F.}~\bibnamefont{{Martini}}},
  \bibnamefont{and} \bibinfo{author}{\bibfnamefont{A.}~\bibnamefont{{Politi}}},
  \bibinfo{journal}{Optics Letters} \textbf{\bibinfo{volume}{43}},
  \bibinfo{pages}{855} (\bibinfo{year}{2018}).

\bibitem[{\citenamefont{{Reimer} et~al.}(2014)\citenamefont{{Reimer},
  {Caspani}, {Clerici}, {Ferrera}, {Kues}, {Peccianti}, {Pasquazi}, {Razzari},
  {Little}, {Chu} et~al.}}]{Reimer2014}
\bibinfo{author}{\bibfnamefont{C.}~\bibnamefont{{Reimer}}},
  \bibinfo{author}{\bibfnamefont{L.}~\bibnamefont{{Caspani}}},
  \bibinfo{author}{\bibfnamefont{M.}~\bibnamefont{{Clerici}}},
  \bibinfo{author}{\bibfnamefont{M.}~\bibnamefont{{Ferrera}}},
  \bibinfo{author}{\bibfnamefont{M.}~\bibnamefont{{Kues}}},
  \bibinfo{author}{\bibfnamefont{M.}~\bibnamefont{{Peccianti}}},
  \bibinfo{author}{\bibfnamefont{A.}~\bibnamefont{{Pasquazi}}},
  \bibinfo{author}{\bibfnamefont{L.}~\bibnamefont{{Razzari}}},
  \bibinfo{author}{\bibfnamefont{B.~E.} \bibnamefont{{Little}}},
  \bibinfo{author}{\bibfnamefont{S.~T.} \bibnamefont{{Chu}}},
  \bibnamefont{et~al.}, \bibinfo{journal}{Optics Express}
  \textbf{\bibinfo{volume}{22}}, \bibinfo{pages}{6535} (\bibinfo{year}{2014}).

\bibitem[{\citenamefont{Dutt et~al.}(2015)\citenamefont{Dutt, Luke,
  Manipatruni, Gaeta, Nussenzveig, and Lipson}}]{Dutt2015}
\bibinfo{author}{\bibfnamefont{A.}~\bibnamefont{Dutt}},
  \bibinfo{author}{\bibfnamefont{K.}~\bibnamefont{Luke}},
  \bibinfo{author}{\bibfnamefont{S.}~\bibnamefont{Manipatruni}},
  \bibinfo{author}{\bibfnamefont{A.~L.} \bibnamefont{Gaeta}},
  \bibinfo{author}{\bibfnamefont{P.}~\bibnamefont{Nussenzveig}},
  \bibnamefont{and} \bibinfo{author}{\bibfnamefont{M.}~\bibnamefont{Lipson}},
  \bibinfo{journal}{Phys. Rev. Applied} \textbf{\bibinfo{volume}{3}},
  \bibinfo{pages}{044005} (\bibinfo{year}{2015}).

\bibitem[{\citenamefont{{Hoff} et~al.}(2015)\citenamefont{{Hoff}, {Nielsen},
  and {Andersen}}}]{Hoff2015}
\bibinfo{author}{\bibfnamefont{U.~B.} \bibnamefont{{Hoff}}},
  \bibinfo{author}{\bibfnamefont{B.~M.} \bibnamefont{{Nielsen}}},
  \bibnamefont{and} \bibinfo{author}{\bibfnamefont{U.~L.}
  \bibnamefont{{Andersen}}}, \bibinfo{journal}{Optics Express}
  \textbf{\bibinfo{volume}{23}}, \bibinfo{pages}{12013} (\bibinfo{year}{2015}).

\bibitem[{\citenamefont{{Shirasaki} and {Haus}}(1990)}]{Shirasaki1990}
\bibinfo{author}{\bibfnamefont{M.}~\bibnamefont{{Shirasaki}}} \bibnamefont{and}
  \bibinfo{author}{\bibfnamefont{H.~A.} \bibnamefont{{Haus}}},
  \bibinfo{journal}{Journal of the Optical Society of America B Optical
  Physics} \textbf{\bibinfo{volume}{7}}, \bibinfo{pages}{30}
  (\bibinfo{year}{1990}).

\bibitem[{\citenamefont{Bergman and Haus}(1991)}]{Bergman1991}
\bibinfo{author}{\bibfnamefont{K.}~\bibnamefont{Bergman}} \bibnamefont{and}
  \bibinfo{author}{\bibfnamefont{H.~A.} \bibnamefont{Haus}},
  \bibinfo{journal}{Opt. Lett.} \textbf{\bibinfo{volume}{16}},
  \bibinfo{pages}{663} (\bibinfo{year}{1991}).

\bibitem[{\citenamefont{{Pfeiffer} et~al.}(2017)\citenamefont{{Pfeiffer},
  {Liu}, {Geiselmann}, and {Kippenberg}}}]{Pfeiffer2017}
\bibinfo{author}{\bibfnamefont{M.~H.~P.} \bibnamefont{{Pfeiffer}}},
  \bibinfo{author}{\bibfnamefont{J.}~\bibnamefont{{Liu}}},
  \bibinfo{author}{\bibfnamefont{M.}~\bibnamefont{{Geiselmann}}},
  \bibnamefont{and} \bibinfo{author}{\bibfnamefont{T.~J.}
  \bibnamefont{{Kippenberg}}}, \bibinfo{journal}{Physical Review Applied}
  \textbf{\bibinfo{volume}{7}}, \bibinfo{eid}{024026} (\bibinfo{year}{2017}).

\bibitem[{\citenamefont{Thomas et~al.}(2018)\citenamefont{Thomas, Dhakal, Raza,
  Peyskens, and Baets}}]{Thomas2018}
\bibinfo{author}{\bibfnamefont{N.~L.} \bibnamefont{Thomas}},
  \bibinfo{author}{\bibfnamefont{A.}~\bibnamefont{Dhakal}},
  \bibinfo{author}{\bibfnamefont{A.}~\bibnamefont{Raza}},
  \bibinfo{author}{\bibfnamefont{F.}~\bibnamefont{Peyskens}}, \bibnamefont{and}
  \bibinfo{author}{\bibfnamefont{R.}~\bibnamefont{Baets}},
  \bibinfo{journal}{Optica} \textbf{\bibinfo{volume}{5}}, \bibinfo{pages}{328}
  (\bibinfo{year}{2018}).

\bibitem[{\citenamefont{{Ralph} and {White}}(1995)}]{Ralph1995}
\bibinfo{author}{\bibfnamefont{T.~C.} \bibnamefont{{Ralph}}} \bibnamefont{and}
  \bibinfo{author}{\bibfnamefont{A.~G.} \bibnamefont{{White}}},
  \bibinfo{journal}{Journal of the Optical Society of America B Optical
  Physics} \textbf{\bibinfo{volume}{12}}, \bibinfo{pages}{833}
  (\bibinfo{year}{1995}).

\bibitem[{\citenamefont{{Wilkes} et~al.}(2016)\citenamefont{{Wilkes}, {Qiang},
  {Wang}, {Santagati}, {Paesani}, {Zhou}, {Miller}, {Marshall}, {Thompson}, and
  {O'Brien}}}]{Wilkes2016}
\bibinfo{author}{\bibfnamefont{C.~M.} \bibnamefont{{Wilkes}}},
  \bibinfo{author}{\bibfnamefont{X.}~\bibnamefont{{Qiang}}},
  \bibinfo{author}{\bibfnamefont{J.}~\bibnamefont{{Wang}}},
  \bibinfo{author}{\bibfnamefont{R.}~\bibnamefont{{Santagati}}},
  \bibinfo{author}{\bibfnamefont{S.}~\bibnamefont{{Paesani}}},
  \bibinfo{author}{\bibfnamefont{X.}~\bibnamefont{{Zhou}}},
  \bibinfo{author}{\bibfnamefont{D.~A.~B.} \bibnamefont{{Miller}}},
  \bibinfo{author}{\bibfnamefont{G.~D.} \bibnamefont{{Marshall}}},
  \bibinfo{author}{\bibfnamefont{M.~G.} \bibnamefont{{Thompson}}},
  \bibnamefont{and} \bibinfo{author}{\bibfnamefont{J.~L.}
  \bibnamefont{{O'Brien}}}, \bibinfo{journal}{Optics Letters}
  \textbf{\bibinfo{volume}{41}}, \bibinfo{pages}{5318} (\bibinfo{year}{2016}).

\bibitem[{\citenamefont{Xuan et~al.}(2016)\citenamefont{Xuan, Liu, Varghese,
  Metcalf, Xue, Wang, Han, Jaramillo-Villegas, Noman, Wang et~al.}}]{Xuan2016}
\bibinfo{author}{\bibfnamefont{Y.}~\bibnamefont{Xuan}},
  \bibinfo{author}{\bibfnamefont{Y.}~\bibnamefont{Liu}},
  \bibinfo{author}{\bibfnamefont{L.~T.} \bibnamefont{Varghese}},
  \bibinfo{author}{\bibfnamefont{A.~J.} \bibnamefont{Metcalf}},
  \bibinfo{author}{\bibfnamefont{X.}~\bibnamefont{Xue}},
  \bibinfo{author}{\bibfnamefont{P.-H.} \bibnamefont{Wang}},
  \bibinfo{author}{\bibfnamefont{K.}~\bibnamefont{Han}},
  \bibinfo{author}{\bibfnamefont{J.~A.} \bibnamefont{Jaramillo-Villegas}},
  \bibinfo{author}{\bibfnamefont{A.~A.} \bibnamefont{Noman}},
  \bibinfo{author}{\bibfnamefont{C.}~\bibnamefont{Wang}}, \bibnamefont{et~al.},
  \bibinfo{journal}{Optica} \textbf{\bibinfo{volume}{3}}, \bibinfo{pages}{1171}
  (\bibinfo{year}{2016}).

\bibitem[{\citenamefont{Ji et~al.}(2017)\citenamefont{Ji, Barbosa, Roberts,
  Dutt, Cardenas, Okawachi, Bryant, Gaeta, and Lipson}}]{Ji2017}
\bibinfo{author}{\bibfnamefont{X.}~\bibnamefont{Ji}},
  \bibinfo{author}{\bibfnamefont{F.~A.~S.} \bibnamefont{Barbosa}},
  \bibinfo{author}{\bibfnamefont{S.~P.} \bibnamefont{Roberts}},
  \bibinfo{author}{\bibfnamefont{A.}~\bibnamefont{Dutt}},
  \bibinfo{author}{\bibfnamefont{J.}~\bibnamefont{Cardenas}},
  \bibinfo{author}{\bibfnamefont{Y.}~\bibnamefont{Okawachi}},
  \bibinfo{author}{\bibfnamefont{A.}~\bibnamefont{Bryant}},
  \bibinfo{author}{\bibfnamefont{A.~L.} \bibnamefont{Gaeta}}, \bibnamefont{and}
  \bibinfo{author}{\bibfnamefont{M.}~\bibnamefont{Lipson}},
  \bibinfo{journal}{Optica} \textbf{\bibinfo{volume}{4}}, \bibinfo{pages}{619}
  (\bibinfo{year}{2017}).

\bibitem[{\citenamefont{Alexander et~al.}(2018)\citenamefont{Alexander,
  Yokoyama, Furusawa, and Menicucci}}]{Alexander2018}
\bibinfo{author}{\bibfnamefont{R.~N.} \bibnamefont{Alexander}},
  \bibinfo{author}{\bibfnamefont{S.}~\bibnamefont{Yokoyama}},
  \bibinfo{author}{\bibfnamefont{A.}~\bibnamefont{Furusawa}}, \bibnamefont{and}
  \bibinfo{author}{\bibfnamefont{N.~C.} \bibnamefont{Menicucci}},
  \bibinfo{journal}{Phys. Rev. A} \textbf{\bibinfo{volume}{97}},
  \bibinfo{pages}{032302} (\bibinfo{year}{2018}).

\bibitem[{\citenamefont{Vaidya et~al.}(2019)\citenamefont{Vaidya, Morrison,
  Helt, Shahrokhshahi, Mahler, Collins, Tan, Lavoie, Repingon, Menotti
  et~al.}}]{Vaidya-inprep}
\bibinfo{author}{\bibfnamefont{V.}~\bibnamefont{Vaidya}},
  \bibinfo{author}{\bibfnamefont{B.}~\bibnamefont{Morrison}},
  \bibinfo{author}{\bibfnamefont{L.}~\bibnamefont{Helt}},
  \bibinfo{author}{\bibfnamefont{R.}~\bibnamefont{Shahrokhshahi}},
  \bibinfo{author}{\bibfnamefont{D.}~\bibnamefont{Mahler}},
  \bibinfo{author}{\bibfnamefont{M.}~\bibnamefont{Collins}},
  \bibinfo{author}{\bibfnamefont{K.}~\bibnamefont{Tan}},
  \bibinfo{author}{\bibfnamefont{J.}~\bibnamefont{Lavoie}},
  \bibinfo{author}{\bibfnamefont{A.}~\bibnamefont{Repingon}},
  \bibinfo{author}{\bibfnamefont{M.}~\bibnamefont{Menotti}},
  \bibnamefont{et~al.}, \bibinfo{journal}{In preparation}
  (\bibinfo{year}{2019}).

\end{thebibliography}

\end{document}